\documentclass[twoside,twocolumn,english,aps,prl,showpacs,superscriptaddress]{revtex4}
\usepackage[latin1]{inputenc}
\usepackage{graphicx}

\makeatletter

\usepackage{babel}
\makeatother
\begin{document}

\newcommand{\pd}{\hat{\psi}^{\dagger}}

\newcommand{\ps}{\hat{\psi}}
 
\newcommand{\ph}{\hat{\phi}}

\newcommand{\kt}[1]{\left|#1\right\rangle }

\title{Relaxation and dephasing in a many-fermion generalization of the
Caldeira-Leggett model}

\author{Florian Marquardt}

\affiliation{Department of Physics, Yale University, New Haven, CT 06520, USA}

\author{D.S. Golubev}

\affiliation{Institut für Theoretische Festkörperphysik, Universität Karlsruhe,
76128 Karlsruhe, Germany }

\affiliation{I. E. Tamm Department of Theoretical Physics, P. N. Lebedev Physics
Institute, 119991 Moscow, Russia }

\date{27.7.2004}

\begin{abstract}
We analyze a model system of fermions in a harmonic oscillator potential
under the influence of a fluctuating force generated by a bath of
harmonic oscillators. This represents an extension of the well-known
Caldeira-Leggett model to the case of many fermions. Using the method
of bosonization, we calculate Green's functions and discuss relaxation
and dephasing of a single extra particle added above the Fermi sea.
We also extend our analysis to a more generic coupling between system
and bath, that results in complete thermalization of the system.
\end{abstract}

\pacs{03.65.Yz, 05.30.Fk, 71.10.Pm}

\maketitle
The interaction between a system and its environment is an important
fundamental issue in quantum mechanics. It is at the basis of relaxation
phenomena (like spontaneous emission), is essential for the measurement
process, and leads to the destruction of interference effects ({}``decoherence''
or {}``dephasing''). In the theory of quantum-dissipative systems\cite{weiss},
there are only few exactly solvable models, most notably the Caldeira-Leggett
model\cite{callegg} of a single particle coupled to a bath of harmonic
oscillators. This is the simplest possible model in which friction
and fluctuations appear. If the particle is free, then this model
can be used to study the quantum analogue of Brownian motion. The
model remains exactly solvable if the particle moves in a parabolic
potential (the damped quantum harmonic oscillator). 

However, in many solid state applications, we actually consider dephasing
of an electron inside a Fermi sea. It is difficult to apply the insights
gained from single-particle calculations in such cases, since the
Pauli principle may play an important role in relaxation processes.
There have been comparatively few detailed studies of quantum-dissipative
many-particle systems. Among them we mention a general discussion
of dephasing in a Luttinger liquid \cite{OpenLuttLiquids}, a study
of fermions coupled to independent baths \cite{indepFermions}, and
a formally exact extension of the Feynman-Vernon influence functional
to fermions \cite{FV-Fermi}. In other cases, the Pauli principle
has been introduced {}``by hand'', by keeping only the thermal part
of the bath spectrum \cite{key-9}. 

In this Letter, we study a natural extension of the Caldeira-Leggett
model to a many-fermion case. The model consists of a sea of fermions
populating the lower energy levels of a harmonic oscillator. We are
interested in the effects that arise when a bath is coupled to this
system via a fluctuating spatially homogeneous force. In contrast
to an analogous system of free fermions \cite{ABring}, the bath leads
to transitions between levels, with strong effects of the Pauli principle.
This model might also prove relevant to the discussion of cold fermionic
atoms in a $1d$ harmonic trap \cite{key-12,wonneberger} under the
influence of fluctuations of the trapping potential. 

\begin{figure}
\begin{center}\includegraphics[%
  width=0.90\columnwidth]{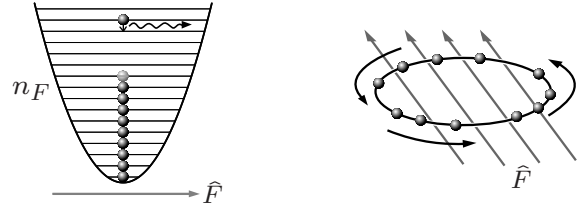}\end{center}

\caption{\label{DFSfigharmon}Left: Fermions in an oscillator, coupled to
a dissipative bath via a fluctuating force $\hat{F}$. Right: Approximately
equivalent model of chiral fermions on a ring, subject to a transverse
force.}
\end{figure}

We rewrite and solve the Hamiltonian using the method of bosonization,
for the case of large particle numbers. This enables us to evaluate
Green's functions and to describe relaxation and dephasing of an extra
particle added above the Fermi sea. Finally, we will extend our model
to a more generic type of coupling.

\emph{The model} - We consider a system of $N$ identical fermions
(non-interacting and spinless) confined in a one-dimensional harmonic
oscillator potential (see Fig. \ref{DFSfigharmon}). A fluctuating
force $\hat{F}$ leads to a coupling of the form $\hat{F}\sum_{j}\hat{x}_{j}$,
yielding, in second quantization:

\begin{eqnarray}
\hat{H}=\omega_{0}\sum_{n=0}^{\infty}n\hat{c}_{n}^{\dagger}\hat{c}_{n}+\hat{H}_{B}+\nonumber \\
\frac{\hat{F}}{\sqrt{2m\omega_{0}}}\sum_{n=0}^{\infty}\sqrt{n+1}(\hat{c}_{n+1}^{\dagger}\hat{c}_{n}+h.c.)\label{DFSham}\end{eqnarray}
The operators $\hat{c}_{n}$ annihilate fermions in the oscillator
levels $n$. The bath Hamiltonian $\hat{H}_{B}$ describes an infinite
number of harmonic oscillators, and the force $\hat{F}$ is a sum
over the bath normal coordinates $\hat{Q}_{j}$. It is characterized
fully by its power spectrum $\left\langle \hat{F}\hat{F}\right\rangle _{\omega}$.
The special case of an Ohmic bath, used for Quantum Brownian motion
\cite{callegg}, has $\left\langle \hat{F}\hat{F}\right\rangle _{\omega}=(\eta\omega/\pi)\theta(\omega_{c}-\omega)\theta(\omega)$
at $T=0$, where $\eta=m\gamma$ is the friction coefficient, $\gamma$
the damping rate, and $\omega_{c}$ the cutoff. As the form of the
coupling (\ref{DFSham}) is not translationally invariant, the frequency
$\omega_{0}$ contains a stabilizing counterterm\cite{callegg,weiss,inpreparation}. 

Effectively, the force acts only on the center-of-mass (c.m.) motion
of the particles which, for the harmonic oscillator, is independent
of the relative motion. Thus, in principle our model reduces to a
single damped harmonic oscillator, analyzed in Ref. \cite{callegg}.
However, we are interested in single-fermion properties, and not in
the collective c.m. motion itself. Although the problem can be solved
exactly via normal modes of the complete set of oscillators and antisymmetrizing
with respect to fermion coordinates, this procedure gets extremely
cumbersome. Instead, we employ an approximation for large fermion
numbers $N$, which also allows an extension to a more generic coupling
between system and bath.

\emph{Bosonization} - For sufficiently large $N$ the lowest levels
are always occupied (at the given interaction strength and temperatures),
i.e. excitations are confined to the region near the Fermi level.
Then we may employ the method of bosonization, rewriting the energy
of fermions as a sum over boson modes \cite{wonneberger}. This is
possible since the energies of the oscillator levels increase linearly
with quantum number, just as the kinetic energy in the Luttinger model
of interacting electrons in one dimension (for recent reviews see
\cite{DelftSchoeller}).

We introduce (approximate) boson operators $\hat{b}_{q}=\frac{1}{\sqrt{q}}\sum_{n=0}^{\infty}\hat{c}_{n}^{\dagger}\hat{c}_{n+q}$
($q\geq1$), which destroy particle-hole excitations. Then, the Hamiltonian
given above becomes approximately

\begin{equation}
\hat{H}\approx\omega_{0}\sum_{q=1}^{\infty}q\hat{b}_{q}^{\dagger}\hat{b}_{q}+\sqrt{\frac{N}{2m\omega_{0}}}\hat{F}(\hat{b}_{1}+\hat{b}_{1}^{\dagger})+\hat{H}_{B}+E_{\hat{N}}\,,\label{DFShapprox}\end{equation}
which will form the basis of our analysis. Here $E_{\hat{N}}=\omega_{0}\hat{N}(\hat{N}-1)/2$
is the total energy of the $N$-fermion noninteracting ground state.
Eq. (\ref{DFShapprox}) reveals that $\hat{F}$ only couples to the
lowest boson mode ($q=1$), corresponding to the c.m. motion. The
damped motion of the c.m. oscillator can be solved exactly, along
the lines of Ref. \cite{callegg} or \cite{key-10}, providing us
with correlators such as $\left\langle \hat{b}_{1}(t)\hat{b}_{1}^{\dagger}(0)\right\rangle $.

\emph{Derivation of Green's functions} - In order to find the Green's
functions, we have to go back from the boson operators $\hat{b}_{q}$
to the fermion operators $\hat{c}_{n}$, by employing well-known finite-size
bosonization identities. In our case, we first have to introduce auxiliary
fermion operators $\hat{\psi}(x)$:

\begin{equation}
\hat{\psi}(x)=\frac{1}{\sqrt{2\pi}}\sum_{n}e^{inx}\hat{c}_{n},\,\hat{c}_{n}=\frac{1}{\sqrt{2\pi}}\int_{0}^{2\pi}e^{-inx}\hat{\psi}(x)\, dx\label{eq:DFScpsi}\end{equation}
The coordinate $x$ does \emph{not} refer to the motion in the oscillator.
Rather, we have effectively mapped our problem to a chiral Luttinger
liquid on a ring with a coupling $\propto\hat{F}\cos(x)$ ($x\in[0,2\pi[$),
see Fig. \ref{DFSfigharmon} (right). Thus, the following results
also describe relaxation of momentum states in that model. Although
a generic discussion of dissipative Luttinger liquids has been provided
in \cite{OpenLuttLiquids}, the particular questions we are going
to study have not been analyzed before. 

The operators $\hat{\psi}(x)$ may be expressed as\cite{DelftSchoeller}:

\begin{equation}
\hat{\psi}(x)=\hat{K}\hat{\lambda}(x)e^{i\hat{\varphi}^{\dagger}(x)}e^{i\hat{\varphi}(x)}=\hat{K}\hat{\lambda}e^{i\hat{\phi}}r\,,\label{DFSpsibos}\end{equation}
with

\begin{equation}
\hat{\phi}=\hat{\varphi}+\hat{\varphi}^{\dagger},\,\,\hat{\varphi}(x)=-i\sum_{q=1}^{\infty}\frac{1}{\sqrt{q}}e^{iqx}\hat{b}_{q}\,.\label{eq:DFSphidef}\end{equation}
The {}``Klein factor'' $\hat{K}$ annihilates a particle, with $[\hat{K},\hat{b}_{q}^{(\dagger)}]=0$
and $\hat{K}(t)=\hat{K}\exp(-i\omega_{0}(\hat{N}-1)t)$. We have $\hat{\lambda}(x)=\exp(i(\hat{N}-1)x)/\sqrt{2\pi}$
and $r\equiv\exp(-[\hat{\varphi}^{\dagger},\hat{\varphi}]/2)$. (The
exponent in $r$ diverges, so a formal cutoff at high $q$ should
be introduced, which will drop out in the end result)

Using Eq. (\ref{eq:DFScpsi}), we find for the hole-propagator:

\begin{equation}
\left\langle \hat{c}_{n'}^{\dagger}(t)\hat{c}_{n}\right\rangle =\frac{1}{2\pi}\int_{0}^{2\pi}e^{i(n'x'-nx)}\left\langle \hat{\psi}^{\dagger}(x',t)\hat{\psi}(x,0)\right\rangle \, dx\, dx'\,.\label{DFSctc}\end{equation}
The $\hat{\psi}$-Green's function is given directly in terms of the
$\hat{\phi}$-correlator, using Eq. (\ref{DFSpsibos}):

\begin{equation}
\left\langle \pd(x',t)\ps(x,0)\right\rangle =\frac{r^{2}}{2\pi}e^{in_{F}((x-x')+\omega_{0}t)}\left\langle e^{-i\hat{\phi}(x',t)}e^{i\hat{\phi}(x,0)}\right\rangle \label{DFSpsigr}\end{equation}
(with $n_{F}=N-1$). The expectation value on the right-hand side
may be evaluated exactly \cite{DelftSchoeller}, since the system-bath
coupling is bilinear. This yields $\exp(E)$ with:

\begin{equation}
E=-\frac{1}{2}\left(\left\langle \hat{\phi}(x',t)^{2}\right\rangle +\left\langle \ph(x,0)^{2}\right\rangle \right)+\left\langle \ph(x',t)\ph(x,0)\right\rangle \label{DFSexpphi}\end{equation}
The correlator of $\ph$ is a polynomial in $X\equiv\exp(ix)$ and
$X'\equiv\exp(ix')$ (see Eq. (\ref{eq:DFSphidef})). Now the double
Fourier integral in Eq. (\ref{DFSctc}) may be evaluated by expanding
$\exp(E)$ as a series in $X$ and $X'$. We find that $\left\langle \hat{c}_{n'}^{\dagger}(t)\hat{c}_{n}\right\rangle $
is the coefficient of $X^{n}/X'^{n'}$ in the expansion of

\begin{equation}
e^{\delta E(X,X',t)}\sum_{k\leq n_{F}}e^{i\omega_{0}kt}(X/X')^{k}\,,\label{expansionExpression}\end{equation}
where the noninteracting exponent has been subtracted in $\delta E=E-E_{(0)}$,
which thus contains only the correlator of the damped c.m. mode $q=1$.
Detailed plots of the Green's function will be published elsewhere
\cite{inpreparation}. Here we provide the result in the weak-coupling
approximation, where we neglect the bath-induced smearing of the equilibrium
Fermi level and use an exponential decay for the c.m. motion. Both
assumptions are summarized in $\left\langle \hat{b}_{1}(t)\hat{b}_{1}\right\rangle =\left\langle \hat{b}_{1}^{\dagger}(t)\hat{b}_{1}\right\rangle =0$
and $\left\langle \hat{b}_{1}(t)\hat{b}_{1}^{\dagger}(0)\right\rangle =e^{-i\omega'_{0}t-\Gamma t/2}$
($\omega'_{0}$ is the renormalized c.m. frequency, and $\Gamma$
is the decay rate, with $\Gamma=N\gamma$ for the Ohmic bath). Then,
the exact Eq. (\ref{expansionExpression}) yields

\begin{equation}
\left\langle \hat{c}_{n}^{\dagger}(t)\hat{c}_{n}\right\rangle \approx e^{i\omega_{0}nt}\sum_{m=0}^{n_{F}-n}\frac{\nu(t)^{m}}{m!}\,,\label{weakcouplGF}\end{equation}
 where $\nu(t)\equiv\exp(-i(\omega_{0}^{'}-\omega_{0})t-\Gamma t/2)-1$.

We find that the hole (particle) propagator does not decay to zero
in the limit $t\rightarrow\infty$, for any $n<n_{F}-1$ ($n>n_{F}+2$),
since $\nu(t)\rightarrow-1$. This is in contrast to the naive single-particle
picture of complete decay for any level $n\neq n_{F},n_{F}+1$ not
directly at the Fermi level (i.e. the result suggested by the leading
order self-energy). Physically, adding a hole (particle) creates an
excited many-particle state which also contains contributions where
the c.m. mode is not excited, and these will not decay, because only
the c.m. mode is damped. A more generic coupling, leading to ergodicity,
will be discussed at the end of this Letter.

\emph{Time-evolution of density matrix -} We now turn to the two-particle
Green's function in order to learn about relaxation of level populations
and dephasing. Consider placing an electron in a superposition of
levels above the Fermi sea, creating the many-particle state $\sum_{n_{0}}\Psi_{n_{0}}\hat{c}_{n_{0}}^{\dagger}\kt{FS}$
at time $0$. We assume the levels $n_{0}$ to be unoccupied. This
will hold for $n_{0}>n_{F}$ in the weak-coupling limit, for which
the following results have been evaluated. The reduced single-particle
density matrix evolves according to:

\begin{equation}
\rho_{nn'}(t)=\sum_{n_{0},n_{0}'}\Psi_{n_{0}}\Psi_{n_{0}'}^{*}\left\langle \hat{c}_{n_{0}'}\hat{c}_{n'}^{\dagger}(t)\hat{c}_{n}(t)\hat{c}_{n_{0}}^{\dagger}\right\rangle \,.\label{redEvolution}\end{equation}
We may rewrite the Green's function in Eq. (\ref{redEvolution}) in
terms of $\hat{\psi}(x)$ (Eq. (\ref{eq:DFScpsi})), leading to a
four-fold Fourier integral, analogous to Eq. (\ref{DFSctc}). Using
Eq. (\ref{DFSpsibos}), this may be evaluated by a series expansion
in four exponentials $\exp(ix^{(')})$, $\exp(iy^{(')})$, similar
to Eq. (\ref{expansionExpression}). We omit the lengthy general formula
\cite{inpreparation}, but discuss a limiting case below.

\begin{figure}
\begin{center}\includegraphics[%
  height=6cm]{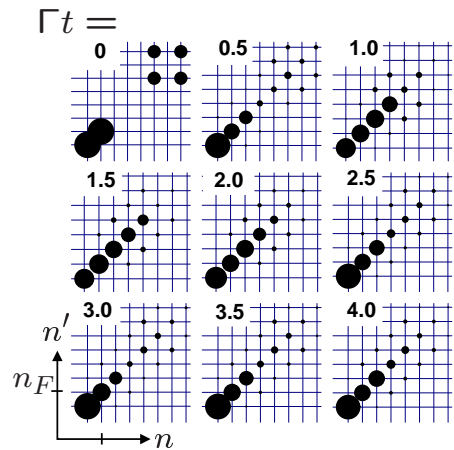}\end{center}

\caption{\label{DFSsuperpositiondecay}Time-evolution of the single-particle
density matrix $\rho_{nn'}(t)$ (Eq. (\ref{redEvolution})), after
placing an extra particle in a superposition of two states. The radius
of each circle gives $\left|\rho_{nn'}(t)\right|$. Levels $n,n'=n_{F}-1,\ldots,n_{F}+6$
are indicated by grid lines. For this example $\omega_{0}-\omega_{0}'=2\Gamma$.}
\end{figure}

In Fig. \ref{DFSsuperpositiondecay}, we have plotted the resulting
time-evolution of the density matrix for the case of an equal superposition
of two levels, $\Psi_{n_{1}}=\Psi_{n_{2}}=1/\sqrt{2}$, at $T=0$.
 The population of the highest occupied states in the Fermi sea (lower
left of panels) decreases, because these fermions become partly excited
at the expense of the extra particle, due to the effective interaction
mediated by the bath. Moreover, the particle does not decay all the
way down to the lowest unoccupied state $n_{F}+1$. Rather, in the
long-time limit, the excitation is distributed over a range of levels
above the Fermi level, up to the initial levels $n_{1,2}$. Again,
this is because only the c.m. mode couples to the bath, such that
a fraction of the initial excitation energy remains in the system.
The same is true of the coherences, i.e. the off-diagonal elements
in the density matrix. 

\begin{figure}
\includegraphics[%
  width=0.99\columnwidth]{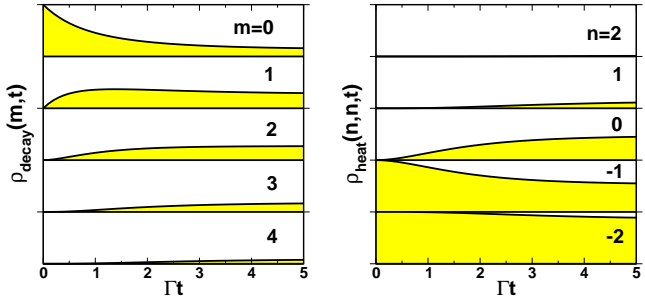}

\caption{\label{LimitingCase}Generic time evolution of density matrix in
the limit of high initial excitation (Eqs. (\ref{rhoSimple})-(\ref{rhoheat});
plot for $\omega_{0}'=\omega_{0}$).}
\end{figure}

\emph{High initial excitation -} These generic features can be analyzed
in more detail for the case of a high initial excitation energy ($n_{0}-n_{F}\gg1$).
Then, the expression for $\rho$ can be given explicitly in a relatively
simple form:

\begin{eqnarray}
 &  & \rho_{nn'}(t)=\sum_{n_{0},n_{0}'}\Psi_{n_{0}}\Psi_{n_{0}'}^{*}e^{i\omega_{0}(n_{0}'-n_{0})t}\delta_{n_{0}-n,n_{0}'-n'}\times\nonumber \\
 &  & [\rho_{{\rm decay}}(n_{0}-n,t)+\rho_{{\rm heat}}(n-n_{F}-1,n'-n_{F}-1,t)]\label{rhoSimple}\end{eqnarray}
The decay of the excitation is described by 

\begin{equation}
\rho_{{\rm decay}}(m,t)=\frac{(-1)^{m}}{m!}\left(\nu(t)+\nu^{*}(t)\right)^{m}e^{\nu(t)+\nu^{*}(t)}\,,\label{rhodecay}\end{equation}
where $m=n_{0}-n=n_{0}'-n'$ may be interpreted as the net number
of quanta transferred to the bath (Fig. \ref{LimitingCase}, left).
At short times, $\Gamma t\ll1$, the nonvanishing entries are $\rho_{{\rm decay}}(1,t)\approx\Gamma t$
and $\rho_{{\rm decay}}(0,t)\approx1-\Gamma t$, i.e. Golden Rule
behaviour is recovered (both for relaxation, $n_{0}=n_{0}'$, and
dephasing, $n_{0}\neq n_{0}'$). In the long-time limit we get a stationary
distribution, $\rho_{{\rm decay}}(m,t)\rightarrow(2^{m}/m!)e^{-2}$. 

{}``Heating'' around the Fermi level is encoded in (see Fig. \ref{LimitingCase})

\begin{equation}
\rho_{{\rm heat}}(n,n',t)=\nu(t)^{n-n'}\sum\frac{\left|\nu(t)\right|^{2(\tilde{m}_{1}+\tilde{m}_{2})}}{m_{1}!m_{2}!\tilde{m}_{1}!\tilde{m}_{2}!}(-1)^{m_{2}+\tilde{m}_{2}}\,,\label{rhoheat}\end{equation}
where the triple sum runs over $\tilde{m}_{1}={\rm max}(0,n'+1)\ldots\infty,$
$\tilde{m}_{2}=0\ldots\infty$, $m_{1}={\rm max}(0,\tilde{m}_{2}+n+1)\ldots n-n'+\tilde{m}_{1}+\tilde{m}_{2}$,
and we have $m_{2}=\tilde{m}_{1}+\tilde{m}_{2}-m_{1}+n-n'$. In the
short-time limit, $\rho_{{\rm heat}}(n,n,t)$ approximates to $1$
for $n<-1$, to $1-|\nu(t)|^{2}$ for $n=-1$, to $|\nu(t)|^{2}$
for $n=0$, and $0$ for $n>0$ (to $O(|\nu|^{2})$), describing the
unperturbed Fermi sea and the onset of heating. Comparing to the full
results (Fig. \ref{DFSsuperpositiondecay}), we find that the limiting
case (\ref{rhoSimple}) is a very good approximation even for small
excitation energies.

\emph{Generic coupling} - Up to now, we have considered a coupling
where the particle coordinates enter linearly, and consequently only
the c.m. mode is damped. 

We now extend our analysis to a more generic situation, replacing
the interaction in Eq. (\ref{DFShapprox}) by: 

\begin{equation}
\sqrt{\frac{N}{2m\omega_{0}}}\hat{F}\sum_{q=1}^{\infty}f_{q}\sqrt{q}(\hat{b}_{q}+\hat{b}_{q}^{\dagger})\,.\label{eq:DFSnewcoupling}\end{equation}
Now the bath induces transitions between levels $n+q$ and $n$, with
an arbitrary (real-valued) amplitude $\propto f_{q}$ (which, however,
must not depend on $n$). For $f_{1}=1,\, f_{q}=0\,(q>1)$ we recover
the original model. For $f_{q}\neq0$ all the boson modes are damped
and couple to each other via the bath. Formally, the correlators $\left\langle \hat{b}_{q'}(t)\hat{b}_{q}^{\dagger}\right\rangle $
can be written in terms of the resolvent of the classical problem
of boson oscillators coupled to bath oscillators (\cite{inpreparation},
compare \cite{key-10}). 

\begin{figure}
\begin{center}\includegraphics[%
  width=0.75\columnwidth]{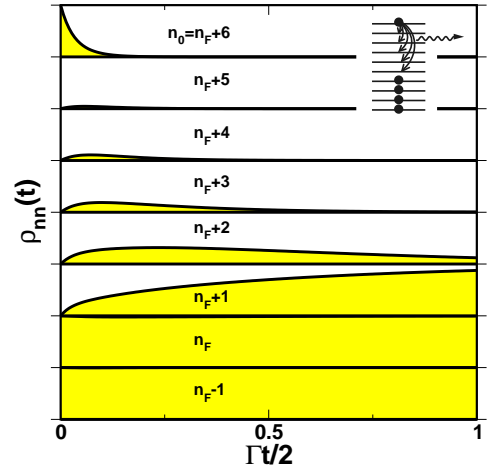}\end{center}

\caption{\label{DFSfigpopalldecay}Decay of populations for a bath inducing
transitions between levels that are arbitrarily far apart (see text). }
\end{figure}

The correlator $\left\langle \ph(x',t)\ph(x,0)\right\rangle $ now
contains contributions for all pairs $q,q'$. The evaluation of the
Green's functions proceeds as before. Unfortunately, one has to deal
with far more terms. However, interesting behaviour is already found
in the weak-coupling limit, which here implies neglecting the effective
coupling between boson modes that has been induced by the bath, and
describing the correlator of each boson mode separately as a damped
oscillation.

For the case of constant $f_{q}=1$ (up to some cutoff) and an Ohmic
bath spectrum, the boson correlator decay rate ($\propto q\left\langle \hat{F}\hat{F}\right\rangle _{\omega=q\omega_{0}}$,
see Eq. (\ref{eq:DFSnewcoupling})) equals $q^{2}\Gamma/2$. This
fits the expectation about Pauli blocking: The decay of a particle
from state $n_{F}+\delta n+1$ is due to transitions by $1$ to $\delta n$
levels, and adding up their rates (which grow linearly) leads to a
total rate $\propto\delta n^{2}$, consistent with the decay rate
of the highest boson mode $q=\delta n$ that is excited by adding
this particle. The actual evolution of the Green's function is a superposition
of decays, with rates up to this value. 

An example for the resulting time-evolution is shown in Fig. \ref{DFSfigpopalldecay}:
Starting from a state where a single extra particle has been added
in level $n_{0}$ above the Fermi sea, one can observe the evolution
of the populations $\rho_{nn}(t)$ (Ohmic bath, $T=0$). At intermediate
times, heating around the Fermi level takes place (barely visible,
in contrast to Fig. \ref{DFSsuperpositiondecay}). In contrast to
the previous case, the relaxation towards the $N+1$-particle ground
state is complete, the system is ergodic. 

\emph{Conclusions} - We have analyzed a many-fermion generalization
of the single particle in a damped harmonic oscillator, illustrating
relaxation and dephasing in a dissipative many-particle system. Using
the method of bosonization (in the limit of large particle number),
we have derived exact expressions for the Green's functions and discussed
them in limiting cases. We have analyzed the decay of an excited state
created by adding one particle above the Fermi level, where one can
observe the {}``heating'' around the Fermi level (due to the effective
interaction between particles), as well as the incomplete decay of
the excited particle. Finally, we have extended our analysis to a
more generic type of coupling between system and bath, where the system
becomes fully ergodic.

\begin{acknowledgments}
We thank C. Bruder, H. Grabert, D. Loss, P. Howell, A. Zaikin, F.
Meier and A.~A. Clerk for comments and discussions. The work of F.M.
has been supported by the Swiss NSF, the NCCR Nanoscience and a DFG
grant.
\end{acknowledgments}


\begin{thebibliography}{10}
\bibitem{weiss}U. Weiss: \emph{Quantum Dissipative Systems}, World Scientific, Singapore
(2000).
\bibitem{callegg}A. O. Caldeira and A. J. Leggett, Physica \textbf{121A}, 587 (1983);
Phys. Rev. A \textbf{31}, 1059 (1985).
\bibitem{OpenLuttLiquids}A. H. C. Neto, C. D. Chamon, C. Nayak, Phys. Rev. Lett. \textbf{79},
4629 (1997).
\bibitem{indepFermions}J. M. Wheatley, Phys. Rev. Lett. \textbf{67}, 1181 (1991). P. C. Howell
and A. J. Schofield, cond-mat/0103191.
\bibitem{FV-Fermi}D. S. Golubev and A. D. Zaikin, Phys. Rev. B \textbf{59}, 9195 (1999).
\bibitem{key-9}B. L. Altshuler, A. G. Aronov, and D. E. Khmelnitsky, J. Phys. C Solid
State \textbf{15}, 7367 (1982); S. Chakravarty and A. Schmid, Phys.
Rep. \textbf{140}, 195 (1986). A. Stern, Y. Aharonov, and Y. Imry,
Phys. Rev. A \textbf{41}, 3436 (1990); D. Cohen and Y. Imry, Phys.
Rev. B \textbf{59}, 11143 (1999).
\bibitem{ABring}F. Marquardt and C. Bruder, Phys. Rev. B \textbf{65}, 125315 (2002).
\bibitem{key-12}F. Schreck \emph{et al.}, Phys. Rev. Lett. 87, 080403 (2001); S. R.
Granade \emph{et al.}, \emph{ibid.} \textbf{88}, 120405 (2002); T.
Loftus \emph{et al.}, \emph{ibid.} \textbf{88}, 173201 (2002); A.
Recati \emph{et al.}, \emph{ibid.} \textbf{90}, 020401 (2003).
\bibitem{wonneberger}W. Wonneberger, Phys. Rev. A \textbf{63}, 063607 (2001); G. Xianlong
and W. Wonneberger, Phys. Rev. A \textbf{65}, 033610 (2002); G. Xianlong,
F. Gleisberg, F. Lochmann, and W. Wonneberger, Phys. Rev. A \textbf{67},
023610 (2003); G. Xianlong and W. Wonneberger, J. Phys. B \textbf{37},
2363 (2004).\bibitem{inpreparation}F. Marquardt and D. S. Golubev, cond-mat/0409401.
\bibitem{DelftSchoeller}J. v. Delft and H. Schoeller, Annalen der Physik, Vol. \textbf{4},
225 (1998). H. Grabert, in \char`\"{}Exotic States in Quantum Nanostructures\char`\"{}
ed. by S. Sarkar, Kluwer (2001).
\bibitem{key-10}V. Hakim and V. Ambegaokar, Phys. Rev. A \textbf{32}, 423 (1985).
\end{thebibliography}
\end{document}